\documentclass[a4paper,11pt,french]{article}
\usepackage[a4paper,includemp=false,body={16cm,24.7cm},vcentering]{geometry}
\usepackage[latin1]{inputenc}
\usepackage[T1]{fontenc}
\usepackage{mathptmx} 
\usepackage[bf]{caption}
\usepackage{graphicx} 
\usepackage{graphics}
\usepackage{subfigure}
\usepackage{picinpar}
\usepackage{picins}
\usepackage{vmargin}
\usepackage{color}
\usepackage{colortbl}
\usepackage{mathptmx}
\usepackage[table]{xcolor}
\usepackage{epsfig}
\usepackage{amsmath}
\usepackage{amssymb}
\usepackage{textcomp}
\usepackage{amsfonts}
\usepackage{longtable}
\usepackage{tabularx}
\usepackage{slashbox}
\usepackage{marvosym}
\usepackage{wasysym}
\usepackage{url}
\usepackage{longtable} 
\usepackage{lscape}
\usepackage{pdfpages}

\usepackage{hyperref,amsmath,amssymb,amsthm}

\usepackage{multirow}

\makeatletter
\renewcommand{\section}{\@startsection{section}{1}{0mm}{30pt}{12pt}{\normalfont\normalsize\bfseries}}

\renewcommand{\subsection}{\@startsection{subsection}{2}{0mm}{18pt}{12pt}{\normalfont\normalsize\itshape}}
\makeatother
\newcommand{\Title}[1]{\begin{center}{\bfseries\fontsize{12pt}{12pt}\selectfont#1}\end{center}}
\newcommand{\Author}[2]{\begin{center}{\fontsize{12pt}{12pt}\selectfont#1}\\{\it #2~}\end{center}}
\newcommand{\Introduction}{\section*{Introduction}}

\begin{document}

\Title{Near-Earth asteroids orbit propagation with Gaia observations}
  
\Author{D. Bancelin$^1$, D. Hestroffer$^1$, W. Thuillot$^1$}{1. IMCCE/Paris observatory, 75014 Paris, France. bancelin@imcce.fr, \\}
 
\Introduction

\noindent

Gaia is an astrometric mission that will be launched in 2013 and set on L2 point of Lagrange. It will observe a large number of Solar System Objets
(SSO) down to magnitude 
20.  The  Solar  System  Science  goal  is  to  map  thousand  of  Main  Belt  Asteroids  (MBAs),  Near  Earth  Objects  (NEOs)  (including  comets) 
and  also  planetary  satellites  with  the 
principal purpuse of orbital determination (better than 5 mas astrometric precision), determination  of asteroid mass, spin properties and taxonomy.
Besides, Gaia will be able 
to discover a few objects, in particular NEOs in the region down to the solar elongation 45$^\circ$  which are harder to detect with current
ground-­based surveys. But Gaia is not a 
follow-­up mission and newly discovered objects can be lost if no ground-­based recovery is processed.  
The purpose of this study is to quantify the impact of Gaia data for the known NEAs population and to show how to handle the problem of these
discoveries when faint number of observations and thus very short arc is
provided.

\section{The Gaia mission}


\noindent
During  the  5-­years  mission, Gaia  will  continously scan the sky  with  a  specific  strategy:  objects  will  be observed  from 
two lines  of sight separated with a constant basic  angle.    The  angle between the Sun direction and the  spin  axis  is  set  to  45$^\circ$. 
The initial  spin  rate  is  1''/min  and the  spin  will  precess  around the  Sun-­Earth  direction  with  a mean period of 63 days.
Because  of  this  specific  scanning  law  and  its  positionning,  Gaia  won't  be  able  to observe down to the solar elongation $\sim$ 45$^\circ$.
But
we do expect some observations and/or discovery of Atira asteroids (moving below the Earth orbit). Two other constants are still free parameters: the
initial spin phase which has an influence on the observation's dates and the initial precession angle which has an influence on the number of
observations for a given target. Because of this specific scanning law, some asteroids can be well-observed -- i.e. the Gaia
observations cover at least one revolution period of the asteroids -- and some others can be poorly-observed -- i.e. the Gaia observations are faint
and cover less than half the revolution period.

\section{Astrometry for known NEAs}

Among the NEAs that will be observed by Gaia, we do expect some observations of Potentially Hazardous Asteroids (PHAs). Those asteroids can show
particular threat of collision with the Earth in the future. To illustrate the impact of Gaia observations on PHAs orbit, we will consider here the
case of the asteroid (99942) Apophis (previously designed 2004 MN$_4$). This asteroid will have a deep close encounter with the Earth in April 2029
within $\sim$ 38000 km and because of the chaoticity of the 2029-post orbit, collisions with the Earth are possible after this date \cite{bancelin12}.

\subsection{Observations of asteroid (99942) Apophis}

\begin{figure}[h!]
\centerline{
   \begin{tabular}{cc}
\includegraphics[width=0.7\columnwidth]{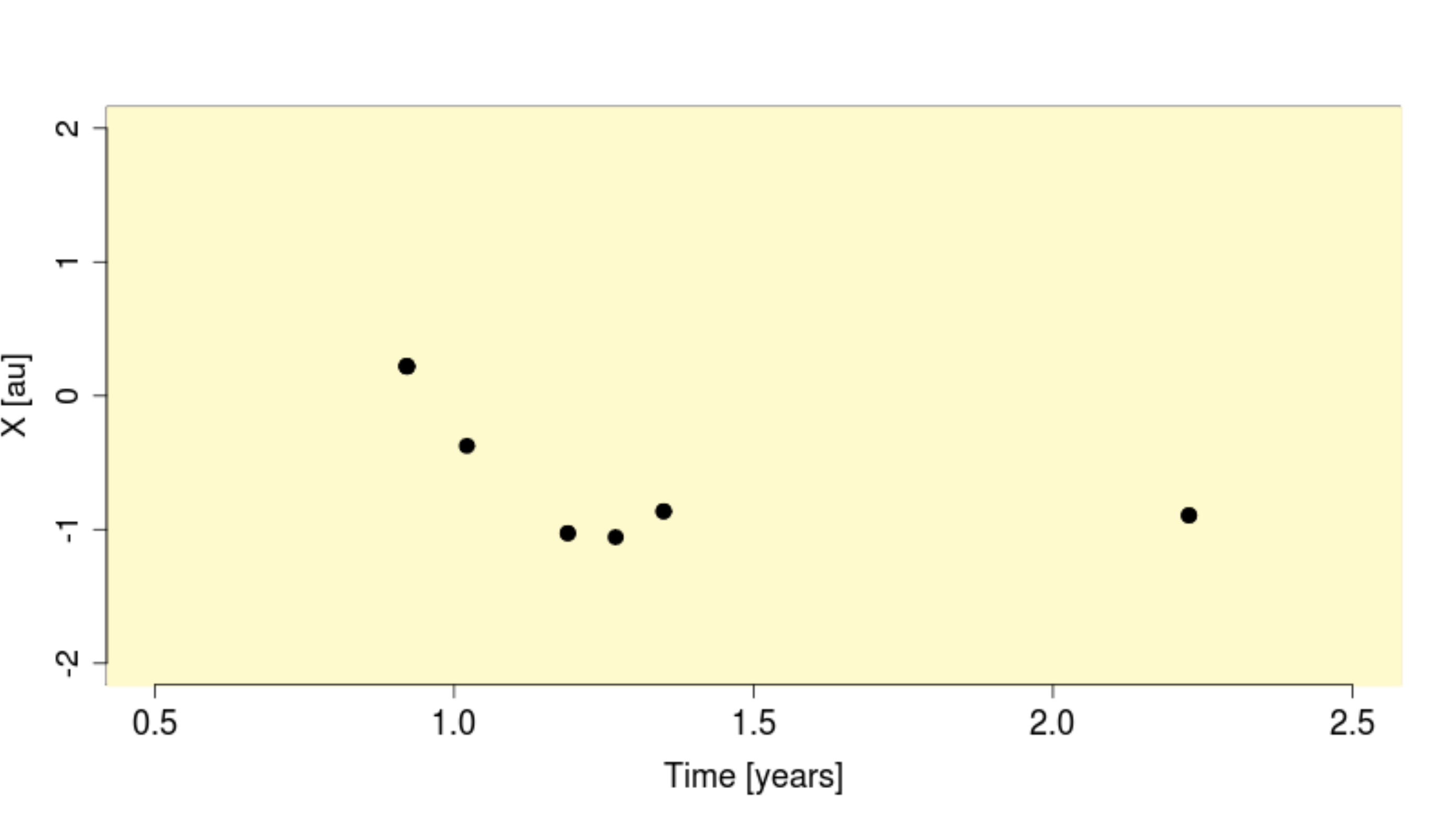} &
\includegraphics[width=0.35\columnwidth]{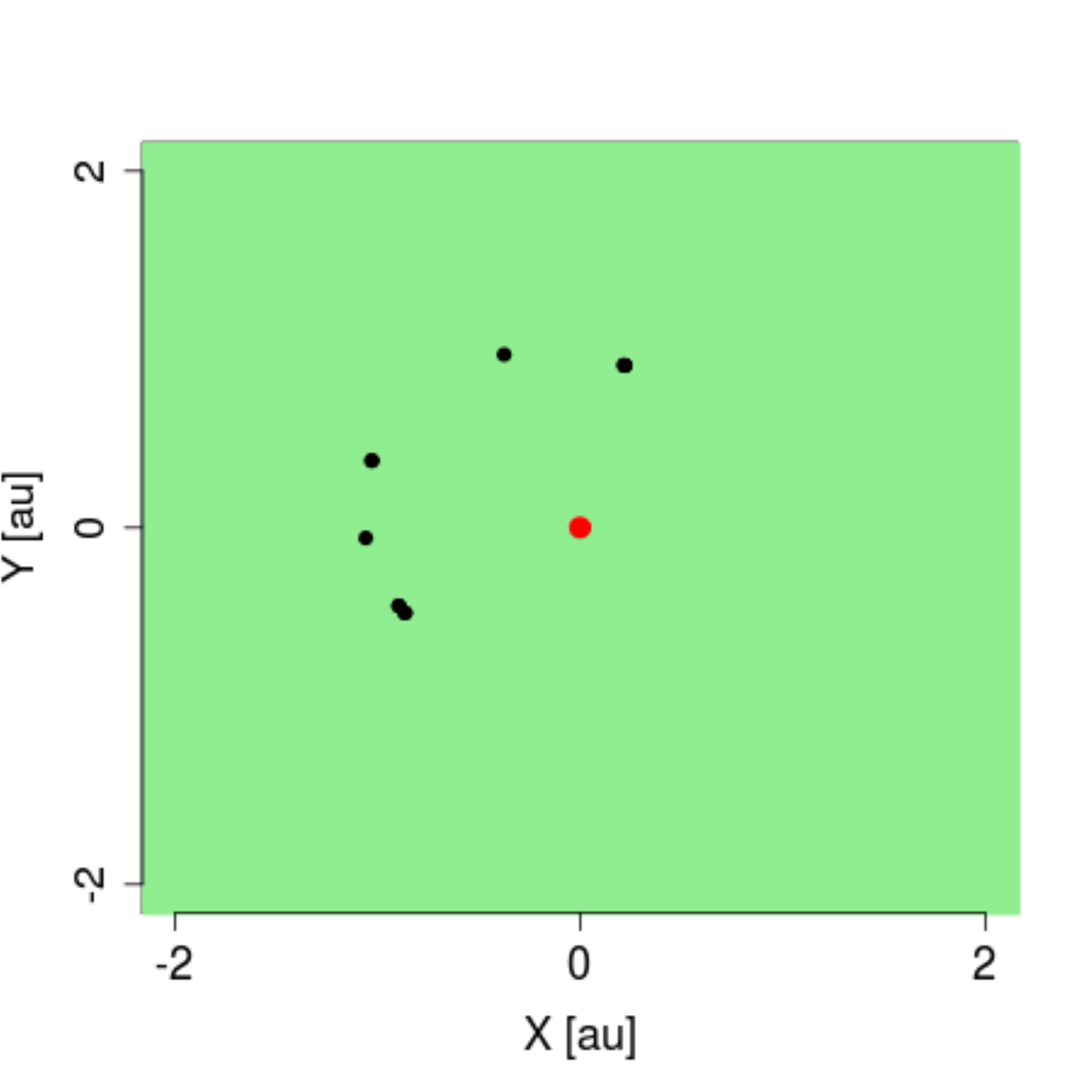}\\
\end{tabular}}
\caption{Left: Gaia observations of Apophis versus time. The x-axis is expressed in terms of the number of years elapsed since
the beginning of the mission. Right: spatial distribution of the observations in the ecliptic frame and centered on the Sun
(\textcolor{red}{$\bullet$}).}\label{F:apophis_gaia}
\end{figure}

Because of the nominal scanning law of Gaia, and in particular the initial precession angle, the number of observations per object can be
inhomogeneous. We can have more than 20 observations as well as less than 10 observations. For our simulations,
we chose a set with the longest arc length (with 12 Gaia observations) and with a 5 mas accuracy. This set covers half the orbit of Apophis (Fig.
\ref{F:apophis_gaia}).

\subsection{Orbital improvement}

\begin{figure}[h!]
\centering
\includegraphics[width=0.7\columnwidth]{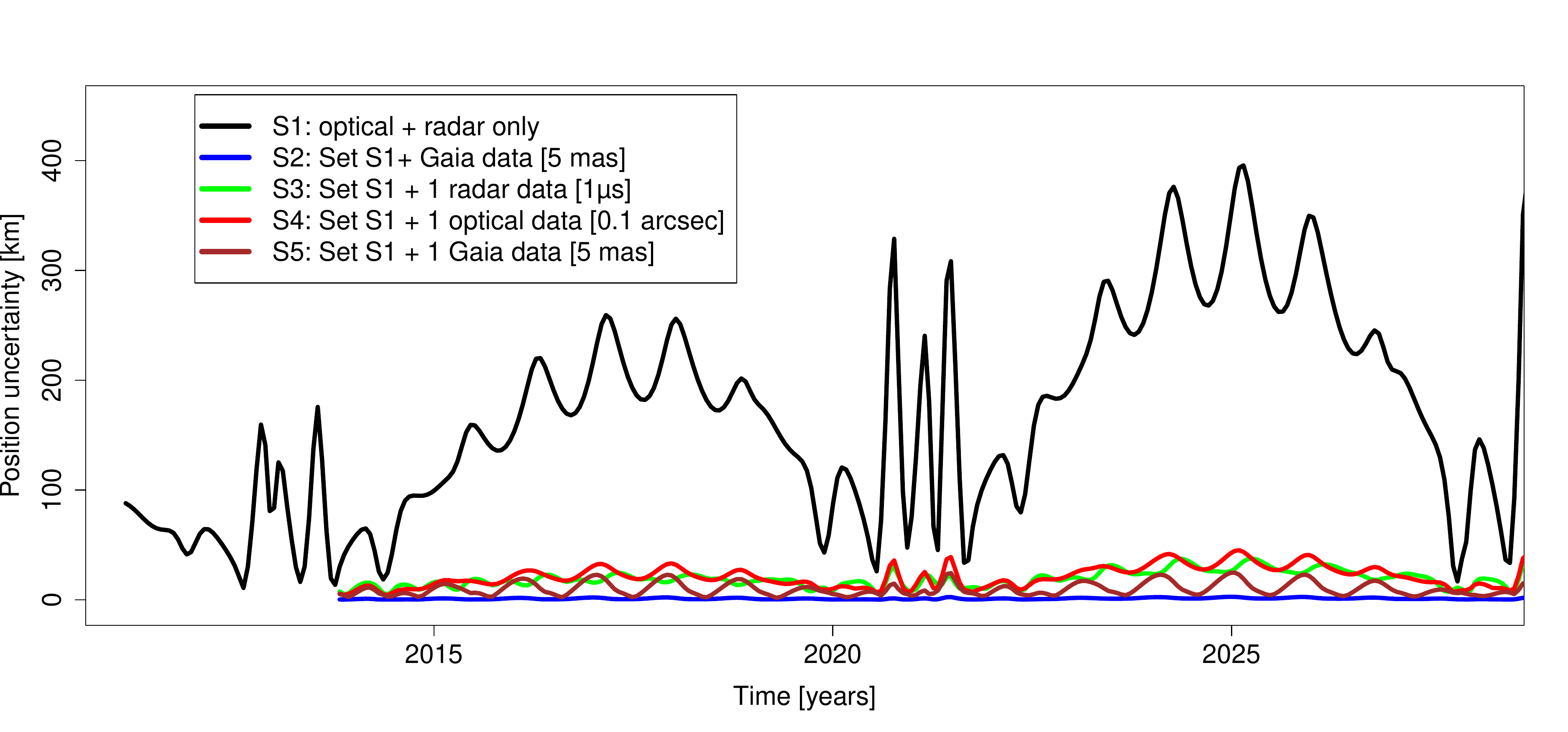}
\caption{Position uncertainty propagation considering various sets of observations. While S$_3$, S$_4$, S$_5$ reduce the
uncertainty to the same level, S$_2$ (using a set of Gaia data) decreases the uncertainty to the kilometer level.}\label{F:uncertainty}
\end{figure}

In the short term, one set of Gaia observations could substantially enhance the current accuracy of the keplerian orbital elements of Apophis (and in
general for all
the possible observed NEAs). Together with all the available ground-based observations (optical and radar), the Gaia observations will enable
to improve the 1$\sigma$ uncertainty of the semi-major by a factor 1000. Besides, the long term uncertainty can be assessed using a linear
propagation of the initial covariance matrix (provided by the least square solution). Comparing various sets of observations (Fig.
\ref{F:uncertainty}) -- each set providing a nominal solution -- one can see that one Gaia data (set S$_5$) is enough to reduce the uncertainty to
the same level as for the sets S$_3$ (with an additional radar data) and S$_4$ (with an additional optical data). But, the impact of one set of Gaia
data is incomparable as the uncertainty is reduced to the kilometer level.

\section{Astrometry for newly discovered asteroids}
When NEAs are discovered, a strategy of recovery can be undertaken. At the epoch of the discovery, Gaia will provide at most two observations
separated by approximately $\Delta$t$ \sim$ 1.5 h. Thus, if it is identifed as an alert, those coordinates will be sent to the Earth within 24 h. But
Gaia is not a follow-up mission and the newly discovered object can be rapidly lost if no ground-based recovery is performed. A follow-up network for
Solar system objects (Gaia-FUN-SSO) has been set-up in order to monitor those asteroids after their discovery \cite{bancelin12b, thuillot11}.\\
In order to optimize the alert mode, we have first to quantify the number of alert expected.

\subsection{Near-Earth asteroids alerts}

\begin{figure}[h!]
\centering
\includegraphics[width=0.7\columnwidth]{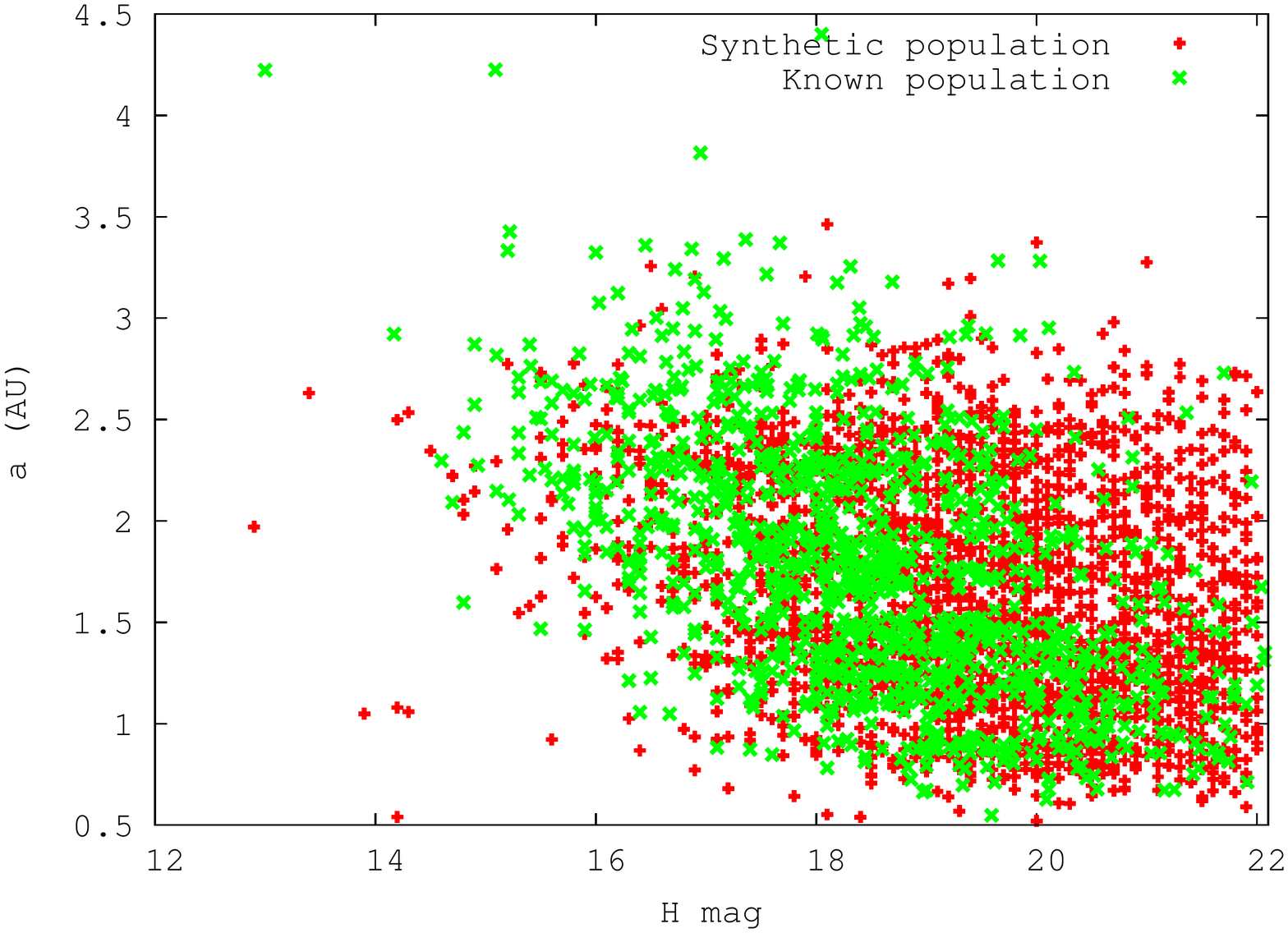}
\caption{Representation of the known (\textcolor{green}{$\times$}) and synthetic (\textcolor{red}{+}) populations possibly observed by the
satellite Gaia during the mission.}\label{F:alerts}
\end{figure}

We presently know more than 9000 NEAs and only $\sim$ 1/6 of this population could be observed by Gaia. For the discovery quantification, we consider
a synthetic population of 30000 NEAs from the model of \cite{greenstreet11}, limited to H $\le$ 22.0. We represented in Fig. \ref{F:alerts} both the
known and synthetic NEAs population that will be observed by Gaia. These populations are represented in the (a,H) space.

\begin{figure}[h!]
\centering
\includegraphics[width=0.7\columnwidth]{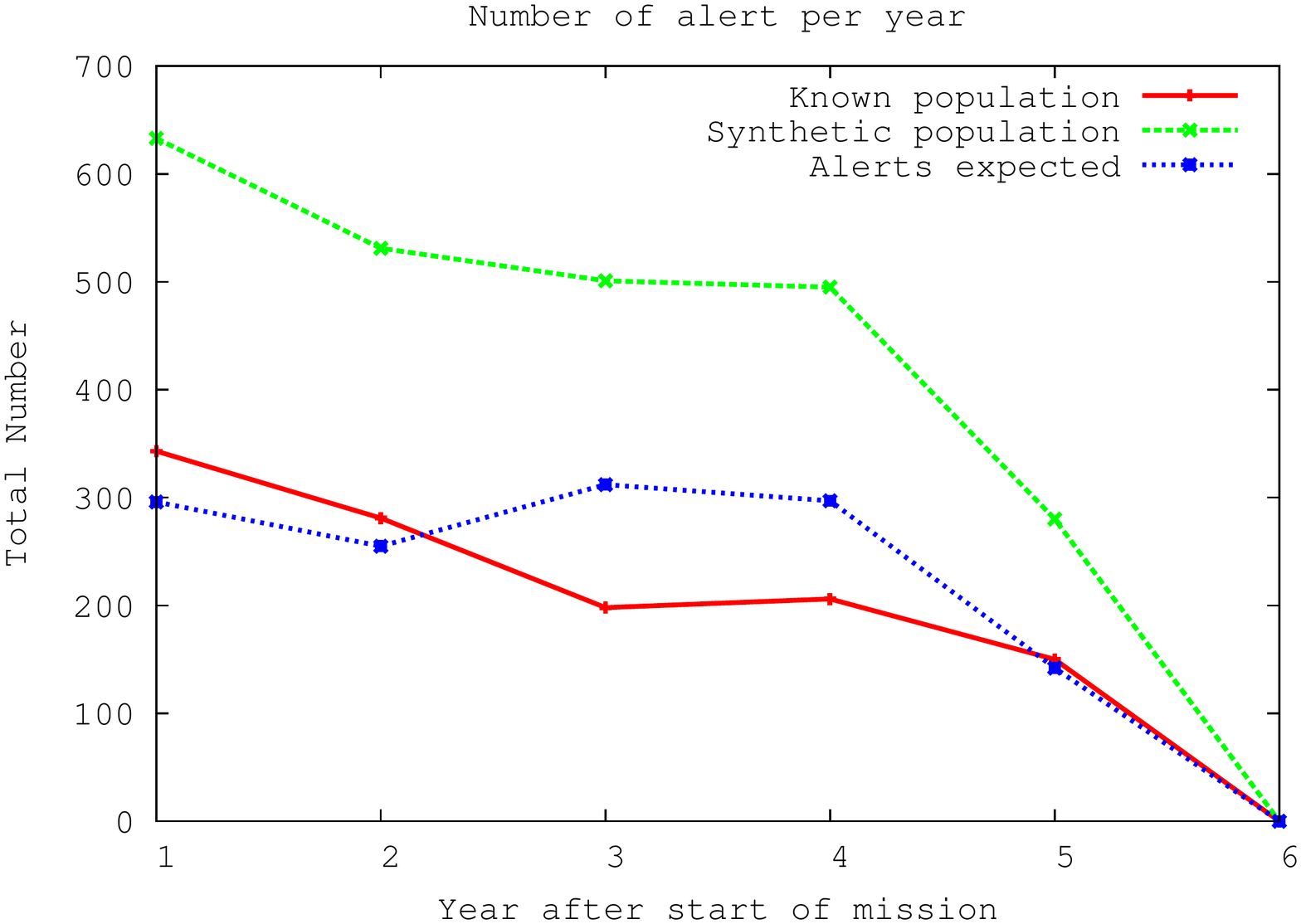}
\caption{Number of alerts (\textcolor{blue}{$\blacksquare$}) compared with the number of observed synthetic NEAs (\textcolor{green}{$\blacksquare$})
and known NEAs (\textcolor{red}{$\blacksquare$}), per year after the beginning of the mission.}\label{F:alert}
\end{figure}

In order to identify and quantify the number of alerts per year after the beginning of the mission, we removed all the synthetic NEAs for which the
semi-major axis \textit{a} and absolute magnitude \textit{H} lie between the minimum and maximum values of (a,H) defining the known NEAs population
observed (see Fig. \ref{F:alerts}). The results are presented in Fig. \ref{F:alert} and show a mean of 4 or 5 alerts per week during the first
4-years after the start of the mission.

\subsection{Strategy of recovery}

\begin{figure}[h!]
\centering
\begin{tabular}{cc}
\includegraphics[width=0.4\columnwidth]{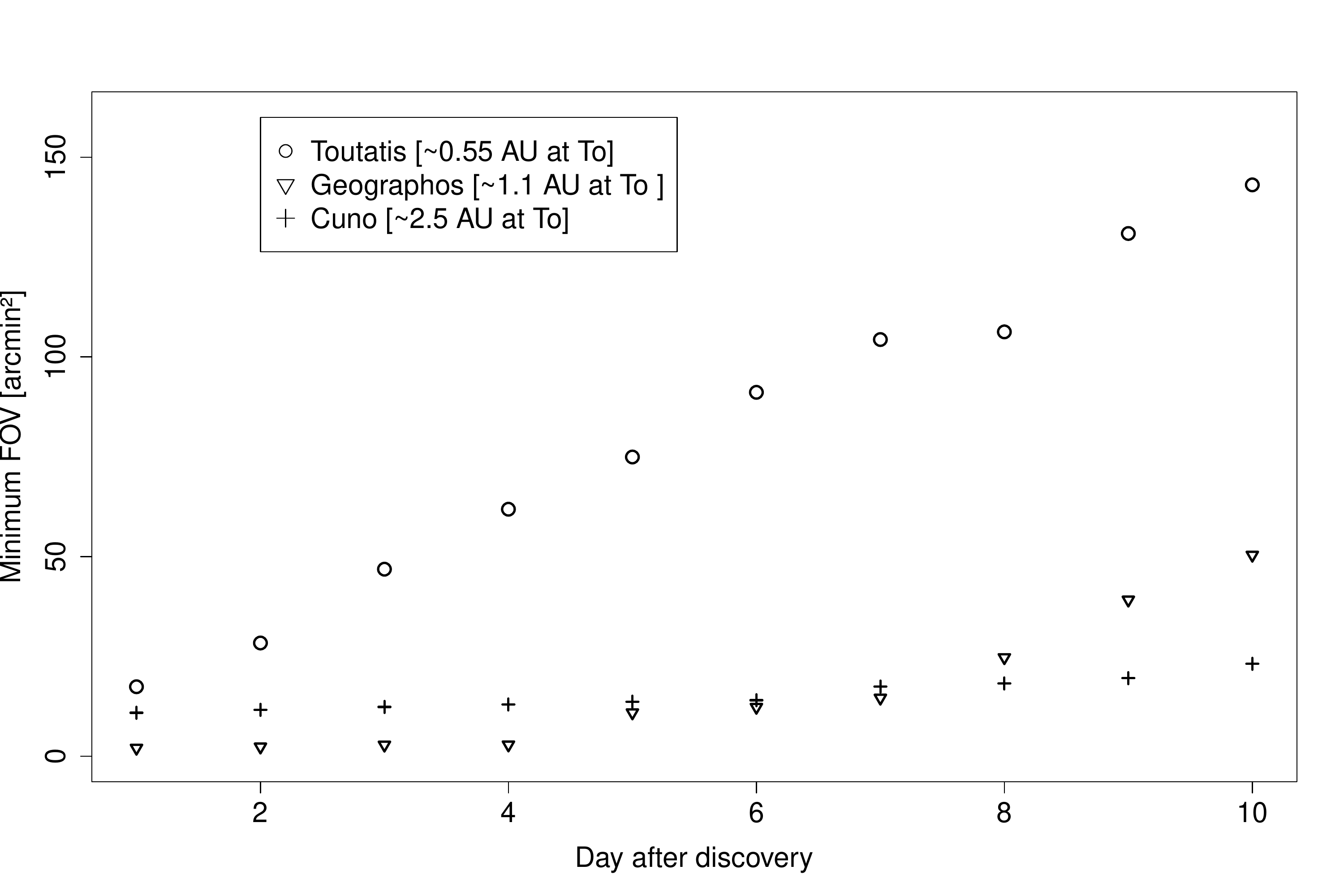} &
\includegraphics[width=0.4\columnwidth]{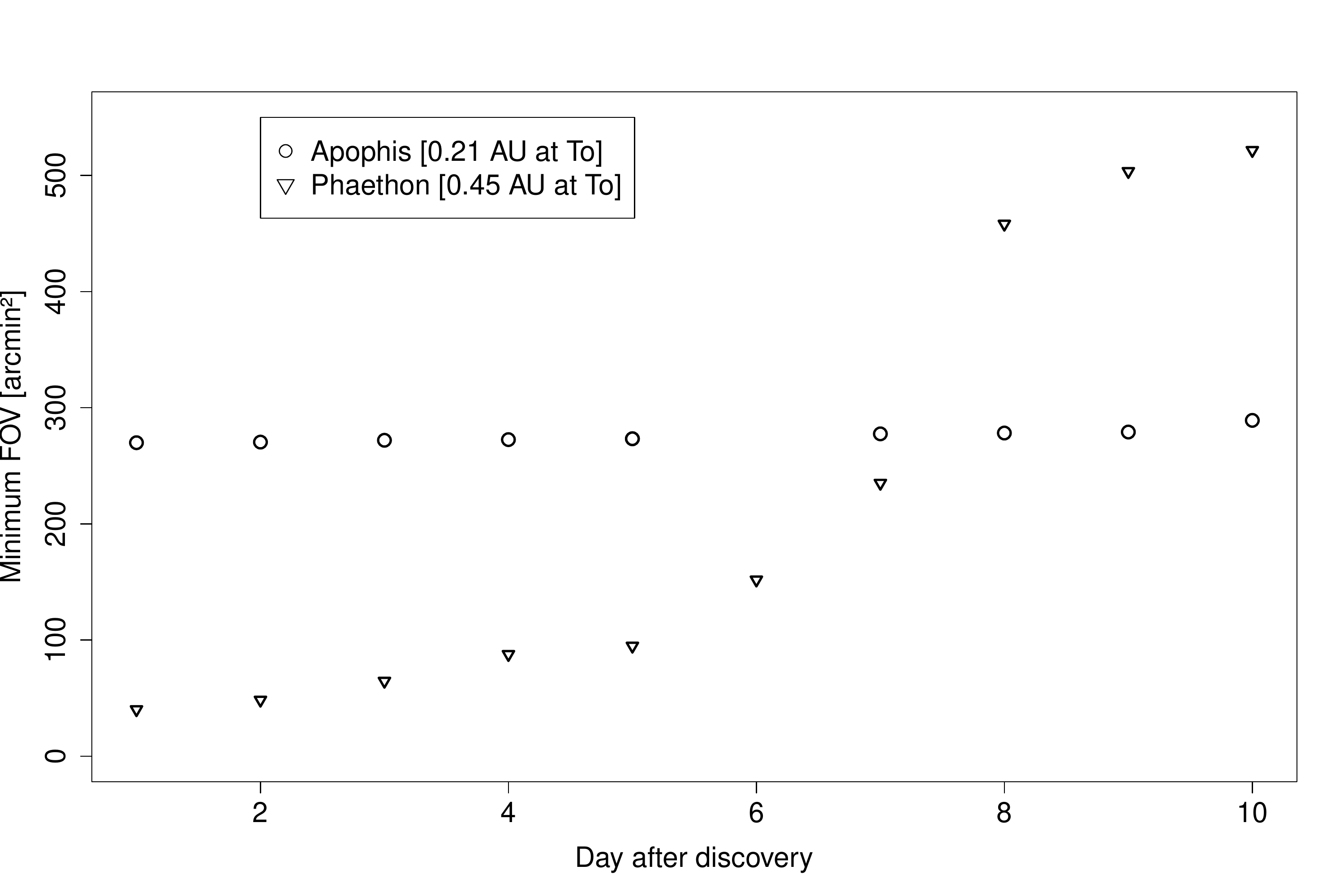}\\
\end{tabular}
\caption{Variation of the minimum FOV required for the recovery process versus time. Left panel: for the asteroids Toutatis, Geographos and Cuno.
Right panel: for the asteroids Apophis and Phaethon.}
\label{F:good_bad}
\end{figure}

When an alert occurs, a preliminary short arc orbit can be computed with the two ($\alpha$, $\delta$) Gaia observations using the Statistical Ranging
method\footnote{this method uses Monte Carlo technique on the ($\alpha$, $\delta$) observations and on the topocentric distances} \cite{virtanen01}.
Thus, a distribution ($\alpha$, $\delta$) can be assessed until a certain number of days after the discovery. Because the distribution can be
quite large, we used statistical tools to extract the maximum likelyhood (ML) of the distribution. Compared to the theoretical position of the object
(given by the orbital elements from astorb database), we can estimate the minimum field of view (FOV) required to recover this object. As shown in
Fig. \ref{F:good_bad}, some asteroids will need typical FOV < 25$\times$25 arcmin$^2$ (case of asteroid Cuno) until 10 days after their discovery,
while some others (case of asteroids Apophis and Phaethon) require a FOV of hundreds of square degrees after their recovery. This behaviour can be
explained by their relative distance to the Earth -- Geographos and Cuno are relatively far from the Earth ($>$ 1 AU) at the epoch of their discovery
by Gaia, and less perturbed by the Earth than the others (distance to the Earth $<$0.5 AU). 

Finally, when the object is recovered by the Gaia-FUN-SSO, complementary ground-based measurements will enable to improve the orbital elements and
the quality of the orbit. This process will enable to optimize the short-term pipeline and the organisation of the network in as
much as, the orbital improvement will enable to use telescopes with smaller FOV and keep the larger ones for asteroids requiring large FOV during the
recovery process.

\section*{Conclusion}

Even  if  Gaia  will  not  be  a  big  NEAs  discoverer,  there  is  a  need  of  the  science 
community  to  support  the  Gaia  mission  in  order  to  be  ready  for  this  opportunity  of 
discovering  new  NEAs.  Among  them,  there  could  be  some  threatening  potentially 
hazardous asteroids and we can not afford to lose them is no Gaia-FUN-­SSO is well­
organized

\end{document}